\documentclass[a4paper,aps,prd,10pt,preprintnumbers,showpacs,twocolumn,superscriptaddress,nofootinbib,nobibnotes,floatfix,amsmath,amssymb]{revtex4-1}
\usepackage{graphicx}
\usepackage{cmap}
\usepackage[utf8]{inputenc}
\usepackage[T1]{fontenc}

\begin{document}

\title{Quasinormal Spectra of Fields of Various Spin in Asymptotically de Sitter Black Holes within Generalized Proca Theory}
\author{Milena Skvortsova}
\email{milenas577@mail.ru}
\affiliation{RUDN University, 6 Miklukho-Maklaya St, Moscow, 117198, Russian Federation}

\begin{abstract}
We study massless scalar, electromagnetic, and Dirac perturbations of asymptotically de Sitter black holes in generalized Proca theory. These geometries are especially interesting because the Proca sector generates both a primary-hair parameter and an effective cosmological term $\Lambda_{\rm eff}$, thereby reshaping the horizon structure and the size of the static patch. Working on this common hairy background, we derive the master equations for the three spin sectors and analyze their quasinormal spectra by means of Pad\'e-improved WKB calculations supplemented by characteristic time-domain integration. We show that the scalar sector, especially the $\ell=0$ mode, is the most sensitive to metric deformations; increasing the Proca-hair parameter $Q$ weakens the damping as the charged three-horizon regime is approached; $\beta$ hardens the spectrum in the $(\alpha,\beta)$ scan; and increasing $\lambda$ and $c_1$ produces the strongest overall softening. For the neutral scalar $\ell=1$ mode, the time-domain Prony extraction agrees excellently with the WKB results and resolves both the Schwarzschild-like black-hole branch and the de Sitter branch. We also discuss the implications of the exact empty-de Sitter limit for strong cosmic censorship and note that the resulting quasinormal frequencies provide useful input for grey-body factors.
\end{abstract}

\maketitle

\section{Introduction}

Quasinormal modes provide one of the most efficient ways to characterize a black-hole geometry through the response of linear perturbations~\cite{Kokkotas:1999bd,Berti:2009kk,Konoplya:2011qq,Bolokhov:2025uxz}. In asymptotically de Sitter spacetimes the problem acquires additional structure because wave propagation is confined to the static region between the event horizon and the cosmological horizon, so the quasinormal spectrum controls not only the ringdown but also the asymptotic decay seen by static observers~\cite{Zhidenko:2003wq,Molina:2003dc,KonoplyaZhidenko:2022Nonosc}. This makes de Sitter black holes especially suitable for a direct comparison between frequency-domain and time-domain approaches. 

A second reason to study quasinormal modes in de Sitter backgrounds is their role in the strong-cosmic-censorship problem for geometries possessing an inner Cauchy horizon~\cite{Cardoso:2017soq,Dias:2018etb,KonoplyaZhidenko:2022SCC}. Even when one is primarily interested in the exterior signal, the least damped modes are the quantities that later enter the spectral-gap diagnostic. For this reason, a systematic spin-by-spin analysis of test fields remains useful even before one turns to the full gravitational sector. Asymptotically de Sitter black holes may also develop various instabilities \cite{Zhu:2014sya,Cuyubamba:2016cug,Konoplya:2013sba}.

Having the above motivations in mind perturbations, stability, scattering and quasinormal modes of asymptotically de Sitter black holes have been studied in nuerous publications \cite{Konoplya:2007zx,Jansen:2017oag,Molina:2003ff,Aragon:2020qdc,Konoplya:2004uk,Zhidenko:2003wq,Konoplya:2017ymp,Mo:2018nnu,Konoplya:2007jv,Kanti:2005xa,Konoplya:2017lhs,Jing:2003wq,Konoplya:2007zx,Dyatlov:2010hq,Konoplya:2014lha}.

The generalized Proca black holes introduced in Refs.~\cite{Charmousis:2025jpx,RefProcaGB2026} are particularly attractive in this context. The same Proca sector that generates the primary hair also fixes the effective cosmological scale of the asymptotics. As a result, one can investigate how the horizon structure and the de Sitter radius are reshaped by the hair parameter while keeping the perturbing fields themselves minimally coupled. A spin comparison is then especially clean: the background is the same, while the differences between the scalar, electromagnetic, and Dirac spectra are driven entirely by the structure of the corresponding effective potentials.

More broadly, the generalized Proca program has been developed well beyond the original action-level construction, with detailed analyses of cosmological dynamics and perturbations, screening around matter sources, and consistent beyond-generalized-Proca extensions in Refs.~\cite{DeFelice:2016CosmoGP,DeFelice:2016ScreenGP,Heisenberg:2016BeyondGP}.

The present manuscript is intended as a dedicated comparative study. Rather than focusing on a massive probe, we consider three massless test fields: a Klein--Gordon scalar, a Maxwell field, and a massless Dirac field. These cases complement one another in a useful way. The scalar sector contains an explicit curvature contribution through the $f'(r)/r$ term in the effective potential, the electromagnetic sector is governed by a purely centrifugal barrier, while the Dirac problem naturally leads to a pair of supersymmetric partner potentials. Studying the three sectors together should therefore clarify which spectral features are universal consequences of the geometry and which are genuinely spin-dependent.

The paper is organized as follows. In Sec.~II we summarize the generalized Proca black-hole background and the relevant asymptotically de Sitter geometry. In Sec.~III we derive the master equations for the massless scalar, electromagnetic, and Dirac fields. Section~IV describes the quasinormal boundary conditions together with the WKB and characteristic time-domain methods to be used in the analysis. In Sec.~V we discuss the main spectral questions that should be addressed in a numerical scan and identify the most informative comparisons between the three spin sectors. Section~VI contains a brief summary.

\section{Generalized Proca Black-Hole Geometry}

We consider the static and spherically symmetric black-hole family obtained in generalized Proca theory~\cite{Heisenberg:2014rta,Charmousis:2025jpx,RefProcaGB2026}. The line element is written as
\begin{equation}
\mathrm{d}s^2=-f(r)\,\mathrm{d}t^2+\frac{\mathrm{d}r^2}{f(r)}+r^2\mathrm{d}\Omega_2^2,
\end{equation}
where the metric function takes the form
\begin{equation}
\begin{split}
 f(r)=&\,1-\frac{2(M-Q)}{r}
 +\frac{r^2}{2\alpha}\Bigg(A \\
 &-\sqrt{B+\frac{8\alpha}{r^3}\left[Q+\frac{\lambda(M-Q)}{2\beta}\right]}\Bigg).
\end{split}
\label{eq:metric_function_massless}
\end{equation}
Here $M$ is the mass-type integration constant, $Q$ is the primary-hair parameter associated with the Proca sector, and
\begin{equation}
A\equiv 1-\frac{\beta\lambda}{2},
\qquad
B\equiv 1-\beta\lambda\left(1-\frac{c_1\lambda}{4}\right).
\end{equation}
For the asymptotically de Sitter branch one has two positive outer roots, the event horizon $r_h$ and the cosmological horizon $r_c$, with the static region located between them.

The asymptotic de Sitter scale can be read off from the large-$r$ behavior of the metric. It is convenient to define
\begin{equation}
\Lambda_{\rm eff}\equiv \frac{3}{2\alpha}\left(\sqrt{B}-A\right),
\qquad
H\equiv \sqrt{\frac{\Lambda_{\rm eff}}{3}},
\end{equation}
so that $H$ plays the role of the effective Hubble parameter. The tortoise coordinate is introduced in the usual way,
\begin{equation}
\frac{\mathrm{d}r_*}{\mathrm{d}r}=\frac{1}{f(r)},
\end{equation}
which maps the static patch onto the full real line, with $r_*\to -\infty$ at the event horizon and $r_*\to +\infty$ at the cosmological horizon.

For the purposes of test-field perturbations, Eq.~\eqref{eq:metric_function_massless} is the only background input required. All spin dependence then enters through the effective potentials of the corresponding master equations.

\section{Massless Test Fields and Their Master Equations}

\subsection{Massless scalar field}

The massless scalar obeys the Klein--Gordon equation
\begin{equation}
\Box \Phi=0.
\end{equation}
With the decomposition \cite{Carter:1968ks,Konoplya:2018arm}
\begin{equation}
\Phi(t,r,\theta,\varphi)=\frac{1}{r}
\sum_{\ell m}\Psi^{(0)}_{\ell m}(t,r)
Y_{\ell m}(\theta,\varphi),
\end{equation}
one obtains the standard wave equation
\begin{equation}
\left(-\frac{\partial^2}{\partial t^2}+\frac{\partial^2}{\partial r_*^2}-V_0(r)\right)
\Psi^{(0)}_{\ell m}(t,r)=0,
\label{eq:scalar_master_massless}
\end{equation}
with effective potential
\begin{equation}
V_0(r)=f(r)\left[\frac{\ell(\ell+1)}{r^2}+\frac{f'(r)}{r}\right].
\label{eq:scalar_potential_massless}
\end{equation}
The extra curvature term $f'(r)/r$ distinguishes the scalar problem from the electromagnetic one and is expected to be most important for low multipoles.

\subsection{Electromagnetic field}

For the Maxwell field,
\begin{equation}
\nabla_\mu F^{\mu\nu}=0,
\qquad
F_{\mu\nu}=\partial_\mu A_\nu-\partial_\nu A_\mu,
\end{equation}
the axial and polar sectors reduce to the same Schr\"odinger-type equation on a static spherically symmetric background. The master variable $\Psi^{(1)}_{\ell m}$ satisfies
\begin{equation}
\left(-\frac{\partial^2}{\partial t^2}+\frac{\partial^2}{\partial r_*^2}-V_1(r)\right)
\Psi^{(1)}_{\ell m}(t,r)=0,
\label{eq:em_master_massless}
\end{equation}
where
\begin{equation}
V_1(r)=f(r)\frac{\ell(\ell+1)}{r^2},
\qquad \ell\geq 1.
\label{eq:em_potential_massless}
\end{equation}
In this sector the barrier is purely centrifugal. Because the same geometry appears in Eq.~\eqref{eq:scalar_potential_massless}, comparing Eqs.~\eqref{eq:scalar_potential_massless} and \eqref{eq:em_potential_massless} offers a particularly direct way to isolate the effect of the curvature term present only for the scalar field.

\subsection{Massless Dirac field}

The massless Dirac equation,
\begin{equation}
\gamma^a e_a{}^{\mu}
\left(\partial_\mu+\Gamma_\mu\right)\Psi=0,
\end{equation}
can be separated on the same background into a pair of one-dimensional wave equations for suitably chosen radial combinations~\cite{Jing:2003wq,LopezOrtega:2007dirac}. Denoting the corresponding master variables by $Z_\pm$, one finds \cite{Brill:1957fx, Cho:2003qe, Jing:2003wq, Kanti:2006ua}
\begin{equation}
\left(-\frac{\partial^2}{\partial t^2}+\frac{\partial^2}{\partial r_*^2}-V_{\pm}(r)\right)
Z_{\pm}(t,r)=0,
\label{eq:dirac_master_massless}
\end{equation}
with supersymmetric partner potentials
\begin{equation}
V_{\pm}(r)=W^2(r)\pm \frac{\mathrm{d}W}{\mathrm{d}r_*},
\qquad
W(r)=\kappa\frac{\sqrt{f(r)}}{r}.
\label{eq:dirac_superpotential_massless}
\end{equation}
Equivalently,
\begin{equation}
V_{\pm}(r)=\kappa^2\frac{f(r)}{r^2}
\pm \kappa\sqrt{f(r)}\left(\frac{f'(r)}{2r}-\frac{f(r)}{r^2}\right),
\end{equation}
where $\kappa=1,2,\ldots$ labels the angular sector. Under the usual quasinormal boundary conditions the two partner potentials are isospectral, so in practice it is sufficient to compute one of them and use the second as a consistency check.

The three effective potentials are therefore all generated by the same function $f(r)$, but they probe it in markedly different ways. This makes the generalized Proca black hole an efficient laboratory for a controlled comparison of spin dependence.

\section{Boundary Conditions and Numerical Methods}

\subsection{Quasinormal boundary conditions}

Quasinormal modes are defined by purely ingoing waves at the event horizon and purely outgoing waves at the cosmological horizon. In the time dependence $e^{-i\omega t}$, this gives
\begin{equation}
\Psi\sim e^{-i\omega(t+r_*)},
\qquad r\to r_h,
\end{equation}
and
\begin{equation}
\Psi\sim e^{-i\omega(t-r_*)},
\qquad r\to r_c.
\end{equation}
The same conditions apply to the scalar, electromagnetic, and Dirac master variables. In the de Sitter setting these boundary conditions isolate the discrete spectrum relevant for the exponentially decaying signal inside the static patch~\cite{Zhidenko:2003wq,Molina:2003dc,KonoplyaZhidenko:2022Nonosc}.

\subsection{Pad\'e-improved WKB approach}

For low-lying modes of single-barrier potentials, the WKB method remains a practical first tool \cite{Iyer:1986np,Konoplya:2003ii,Matyjasek:2017psv,Matyjasek:2019eeu,Matyjasek:2026yiu}. In its standard form the quasinormal frequencies satisfy
\begin{equation}
\frac{i(\omega^2-V_0)}{\sqrt{-2V_0''}}-
\sum_{j=2}^{p}\Lambda_j
=n+\frac{1}{2},
\label{eq:wkb_generic_massless}
\end{equation}
where $V_0$ is the value of the potential at its maximum, $V_0''$ is the second derivative with respect to the tortoise coordinate, $n$ is the overtone number, and $\Lambda_j$ denote higher-order corrections. In practice, the Pad\'e-resummed version of high-order WKB is especially useful because it stabilizes the approximation and provides an internal error estimate through the comparison of neighboring approximants~\cite{Dubinsky:2026wcv,Konoplya:2009hv,Lutfuoglu:2025hjy,Malik:2025erb,Bolokhov:2025aqy,Skvortsova:2024msa,Konoplya:2023ppx,Guo:2020caw,Malik:2025czt,Konoplya:2023moy,Bolokhov:2025lnt,Dubinsky:2025nxv,Eniceicu:2019npi,Lutfuoglu:2025ljm,Kokkotas:2010zd,Malik:2026lfj,Momennia:2018hsm,Dubinsky:2024nzo,Lutfuoglu:2025ohb,Konoplya:2019ppy,Gonzalez:2022ote,Malik:2025ava,Bolokhov:2025egl,Bolokhov:2026kqu}.
In the present problem the WKB method is naturally suited to the oscillatory branch of modes supported by the barrier between the two outer horizons. Its reliability should be assessed case by case, especially for the lowest multipoles and for backgrounds approaching the Nariai regime.

\subsection{Characteristic time-domain integration}

\begin{figure}
\includegraphics[width=0.98\linewidth]{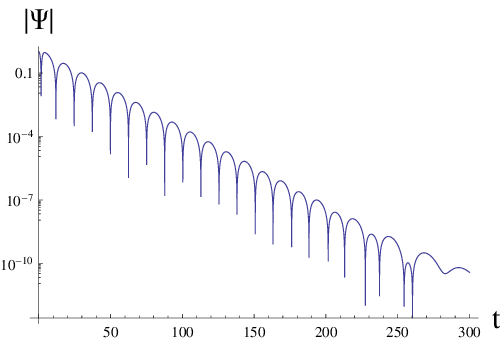}
\refstepcounter{figure}\label{fig:td-reference-neutral-massless}
\par\smallskip
{\small \textbf{FIG.~\thefigure.} Time-domain profile $|\Psi|$ versus $t$ for the scalar $\ell=1$ perturbation of the neutral generalized Proca black-hole background with $(M,Q,\alpha,\beta,\lambda,c_1)=(1,0,1,1,0.2,2)$, shown on a logarithmic scale. A Prony analysis gives $\omega_0=0.249251-0.0849858\,i$ and $\omega_1=0.228648-0.262598\,i$ for the Schwarzschild-like branch, and $\omega_0^{\rm dS}=-0.0525948\,i$ and $\omega_1^{\rm dS}=-0.158721\,i$ for the de Sitter branch. The corresponding Pad\'e-WKB values for the black-hole branch are $0.249251-0.084986\,i$ and $0.228643-0.262594\,i$, while the exact empty-de-Sitter values are $-0.052624\,i$ and $-0.157871\,i$.}
\end{figure}

\begin{table*}[t]
\caption{Fundamental scalar, electromagnetic, and Dirac quasinormal frequencies for the charge scan
$Q=0,0.2,0.4,0.6,0.8,1.0,1.02$ at fixed $M=1$ and $(\alpha,\beta,\lambda,c_1)=(1,1,0.2,2)$.
In each block, $\Delta$ denotes the relative difference between the Pad\'e-improved WKB results at orders 16 and 14.}
\label{tab:qscan_all}
\centering
{\scriptsize
\setlength{\tabcolsep}{3pt}
\renewcommand{\arraystretch}{1.08}
\resizebox{\textwidth}{!}{%
\begin{tabular}{c c c c c c c}
\hline
 & \multicolumn{3}{c}{$\ell=0$ scalar} & \multicolumn{3}{c}{$\ell=1$ scalar} \\
\cline{2-4}\cline{5-7}
$Q$ & WKB16 ($\tilde{m}=8$) & WKB14 ($\tilde{m}=7$) & $\Delta$ & WKB16 ($\tilde{m}=8$) & WKB14 ($\tilde{m}=7$) & $\Delta$ \\
\hline
$0$ & $0.094568-0.094048 i$ & $0.094607-0.093999 i$ & $0.0472\%$ & $0.249251-0.084986 i$ & $0.249251-0.084986 i$ & $0.\times 10^{-4}\%$ \\
$0.2$ & $0.095287-0.093182 i$ & $0.095318-0.093321 i$ & $0.107\%$ & $0.250950-0.084363 i$ & $0.250950-0.084363 i$ & $0\%$ \\
$0.4$ & $0.096710-0.091614 i$ & $0.096709-0.091612 i$ & $0.0015\%$ & $0.254066-0.083032 i$ & $0.254066-0.083032 i$ & $0\%$ \\
$0.6$ & $0.098672-0.088848 i$ & $0.098702-0.088765 i$ & $0.0666\%$ & $0.258882-0.080696 i$ & $0.258882-0.080696 i$ & $0.0002\%$ \\
$0.8$ & $0.100752-0.083686 i$ & $0.100802-0.083670 i$ & $0.0397\%$ & $0.265976-0.076447 i$ & $0.265977-0.076446 i$ & $0.0003\%$ \\
$1.0$ & $0.097558-0.074953 i$ & $0.097526-0.074994 i$ & $0.0422\%$ & $0.275861-0.066904 i$ & $0.275861-0.066904 i$ & $0\%$ \\
$1.02$ & $0.096723-0.074778 i$ & $0.096710-0.074782 i$ & $0.0111\%$ & $0.276824-0.065300 i$ & $0.276824-0.065300 i$ & $0\%$ \\
\hline
\end{tabular}}

\vspace{0.6em}

\resizebox{\textwidth}{!}{%
\begin{tabular}{c c c c c c c}
\hline
 & \multicolumn{3}{c}{Electromagnetic, $\ell=1$} & \multicolumn{3}{c}{Electromagnetic, $\ell=2$} \\
\cline{2-4}\cline{5-7}
$Q$ & WKB16 ($\tilde{m}=8$) & WKB14 ($\tilde{m}=7$) & $\Delta$ & WKB16 ($\tilde{m}=8$) & WKB14 ($\tilde{m}=7$) & $\Delta$ \\
\hline
$0$ & $0.213801-0.079421 i$ & $0.213801-0.079421 i$ & $0\%$ & $0.393271-0.081467 i$ & $0.393271-0.081467 i$ & $0\%$ \\
$0.2$ & $0.215661-0.078919 i$ & $0.215661-0.078919 i$ & $0.\times 10^{-4}\%$ & $0.395972-0.080917 i$ & $0.395972-0.080917 i$ & $0\%$ \\
$0.4$ & $0.219144-0.077826 i$ & $0.219144-0.077826 i$ & $0\%$ & $0.400972-0.079722 i$ & $0.400972-0.079722 i$ & $0\%$ \\
$0.6$ & $0.224628-0.075832 i$ & $0.224628-0.075832 i$ & $0.\times 10^{-4}\%$ & $0.408837-0.077586 i$ & $0.408837-0.077586 i$ & $0\%$ \\
$0.8$ & $0.232989-0.071961 i$ & $0.232989-0.071961 i$ & $0.\times 10^{-4}\%$ & $0.420859-0.073599 i$ & $0.420859-0.073599 i$ & $0\%$ \\
$1.0$ & $0.245779-0.062011 i$ & $0.245779-0.062011 i$ & $0\%$ & $0.440130-0.064043 i$ & $0.440130-0.064043 i$ & $0\%$ \\
$1.02$ & $0.247174-0.060108 i$ & $0.247174-0.060108 i$ & $0.\times 10^{-4}\%$ & $0.442632-0.062259 i$ & $0.442632-0.062259 i$ & $0\%$ \\
\hline
\end{tabular}}

\vspace{0.6em}

\resizebox{\textwidth}{!}{%
\begin{tabular}{c c c c c c c}
\hline
 & \multicolumn{3}{c}{Dirac, $\ell=1/2$} & \multicolumn{3}{c}{Dirac, $\ell=3/2$} \\
\cline{2-4}\cline{5-7}
$Q$ & WKB16 ($\tilde{m}=8$) & WKB14 ($\tilde{m}=7$) & $\Delta$ & WKB16 ($\tilde{m}=8$) & WKB14 ($\tilde{m}=7$) & $\Delta$ \\
\hline
$0$ & $0.157689-0.083104 i$ & $0.157724-0.082855 i$ & $0.141\%$ & $0.326697-0.082533 i$ & $0.326697-0.082533 i$ & $0\%$ \\
$0.2$ & $0.158837-0.082168 i$ & $0.158852-0.082159 i$ & $0.0097\%$ & $0.328765-0.081956 i$ & $0.328765-0.081956 i$ & $0.\times 10^{-4}\%$ \\
$0.4$ & $0.160972-0.080778 i$ & $0.160991-0.080783 i$ & $0.0110\%$ & $0.332589-0.080706 i$ & $0.332589-0.080706 i$ & $0.\times 10^{-4}\%$ \\
$0.6$ & $0.164281-0.078385 i$ & $0.164274-0.078380 i$ & $0.0047\%$ & $0.338591-0.078486 i$ & $0.338591-0.078486 i$ & $0\%$ \\
$0.8$ & $0.169147-0.073999 i$ & $0.169126-0.073996 i$ & $0.0113\%$ & $0.347692-0.074385 i$ & $0.347692-0.074385 i$ & $0.00017\%$ \\
$1.0$ & $0.175060-0.063582 i$ & $0.175054-0.063578 i$ & $0.0037\%$ & $0.361688-0.064790 i$ & $0.361687-0.064789 i$ & $0.\times 10^{-4}\%$ \\
$1.02$ & $0.175229-0.061815 i$ & $0.175228-0.061814 i$ & $0.0011\%$ & $0.363355-0.063058 i$ & $0.363355-0.063057 i$ & $0.\times 10^{-4}\%$ \\
\hline
\end{tabular}}
}
\end{table*}

\begin{figure*}[t]
\centering
\includegraphics[width=0.98\textwidth]{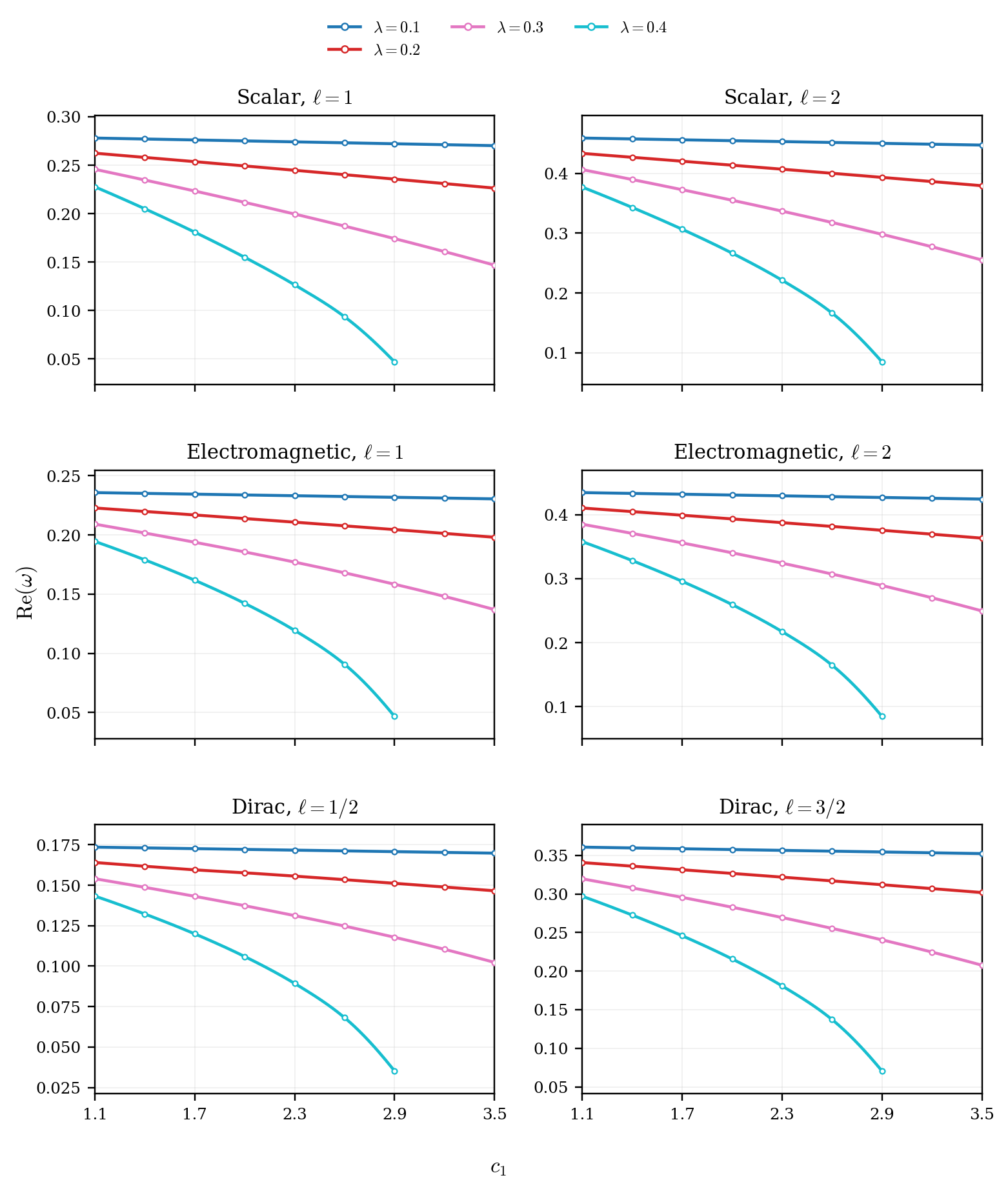}
\caption{Real parts of the fundamental quasinormal frequencies for the $(c_1,\lambda)$ scan of six representative sectors at fixed $Q=0$, $M=1$, and $\alpha=\beta=1$. The scalar panels correspond to $\ell=1$ and $\ell=2$, $c_1$ runs along the horizontal axis, and each curve corresponds to one fixed value $\lambda\in\{0.1,0.2,0.3,0.4\}$. Markers show the tabulated WKB16 points, and the solid lines are interpolations between them.}
\label{fig:lambda_c1_re}
\end{figure*}

\begin{figure*}[t]
\centering
\includegraphics[width=0.98\textwidth]{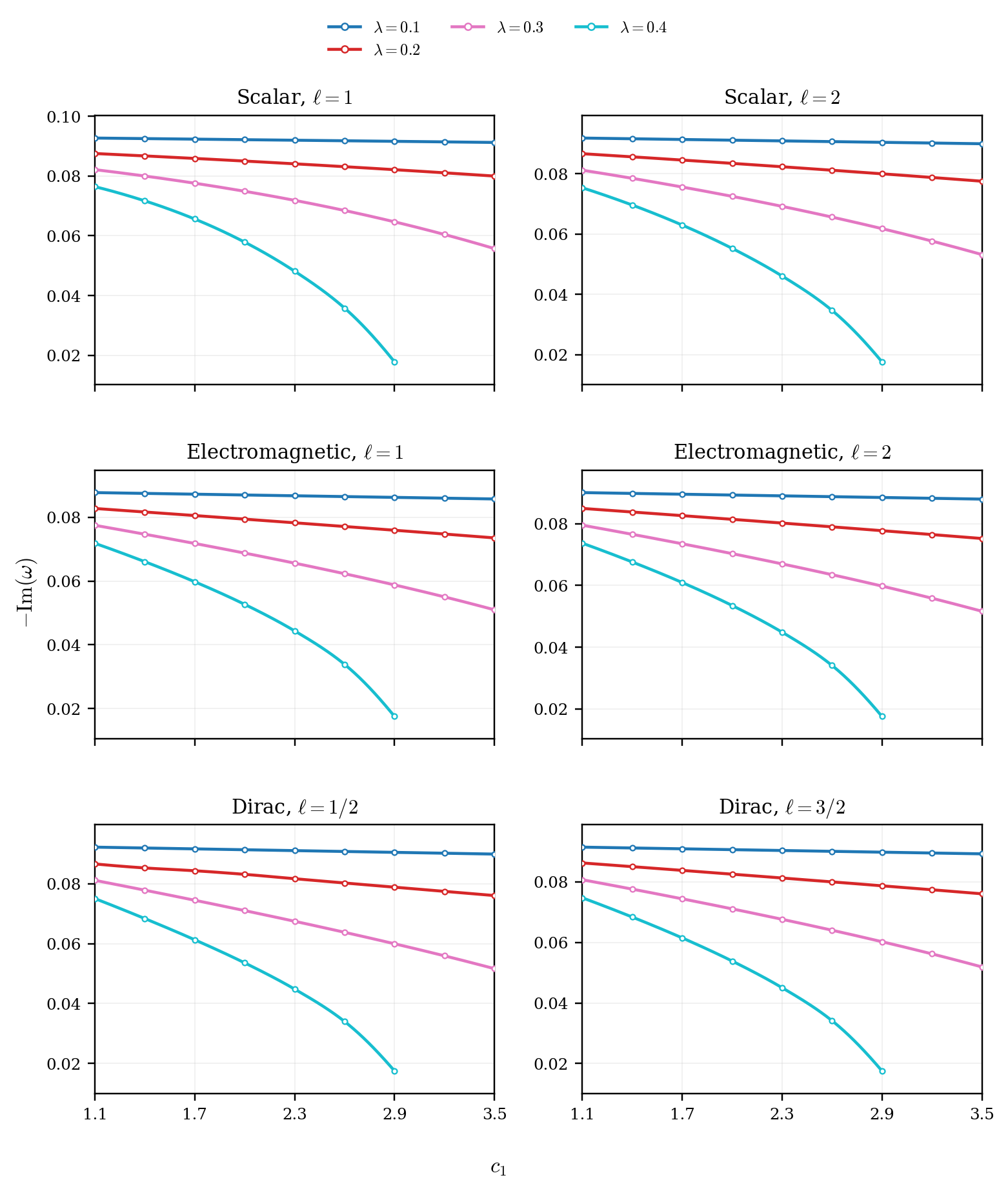}
\caption{Damping rates $-\mathrm{Im}(\omega)$ for the same $(c_1,\lambda)$ scan as in Fig.~\ref{fig:lambda_c1_re}, again plotted as functions of $c_1$ for fixed $\lambda$. The scalar panels correspond to $\ell=1$ and $\ell=2$, while the remaining four panels show the electromagnetic and Dirac sectors. Markers show the tabulated WKB16 points, and the solid lines are interpolations between them.}
\label{fig:lambda_c1_im}
\end{figure*}

The time-domain evolution offers an independent check and, in many cases, a cleaner identification of the dominant late-time signal. Introducing null coordinates
\begin{equation}
 u=t-r_*,
 \qquad
 v=t+r_*,
\end{equation}
the master equation becomes
\begin{equation}
4\frac{\partial^2\Psi}{\partial u\partial v}+V(r)\Psi=0.
\label{eq:null_form_massless}
\end{equation}
On a null grid, the characteristic integration scheme of Gundlach, Price, and Pullin updates the field according to~\cite{GundlachPricePullin:1994,GundlachPricePullin:1994b}
\begin{equation}
\Psi_N=\Psi_W+\Psi_E-\Psi_S
-\frac{\Delta^2}{8}V_S\left(\Psi_W+\Psi_E\right)
+\mathcal{O}(\Delta^4),
\label{eq:gpp_scheme_massless}
\end{equation}
where $(S,E,W,N)$ denote the standard south, east, west, and north points of the diamond cell. This integration scheme was broadly used in numerous works \cite{Konoplya:2023aph,Bolokhov:2023bwm,Skvortsova:2024wly,Malik:2023bxc,Konoplya:2014lha,Arbelaez:2026eaz,Konoplya:2024kih,Qian:2022kaq,Dubinsky:2025fwv,Skvortsova:2024atk,Varghese:2011ku,Bolokhov:2023dxq,Malik:2024nhy,Aneesh:2018hlp,Arbelaez:2025gwj,Abdalla:2012si,Konoplya:2024hfg,Konoplya:2018yrp} and usually showed good agreement with alternative methods.

Once the waveform is generated, the oscillatory segment can be fitted by a Prony expansion in order to extract the first few frequencies directly from the time-domain signal~\cite{Dubinsky:2024gwo}. In asymptotically de Sitter spacetimes this is especially useful because the late-time decay is exponential rather than power-law, so the dominant quasinormal frequencies remain visible deep into the evolution. Indeed, quasinormal modes of three-dimensional asymptotically de Sitter black holes \cite{Konoplya:2020ibi} were computed with high accuracy using time-domain integration in \cite{Skvortsova:2023zca,Skvortsova:2023zmj}, while a similar analysis for the four-dimensional case was carried out in \cite{Dubinsky:2024gwo,Bolokhov:2024ixe}.

\begin{figure*}[t]
\centering
\includegraphics[width=0.98\textwidth]{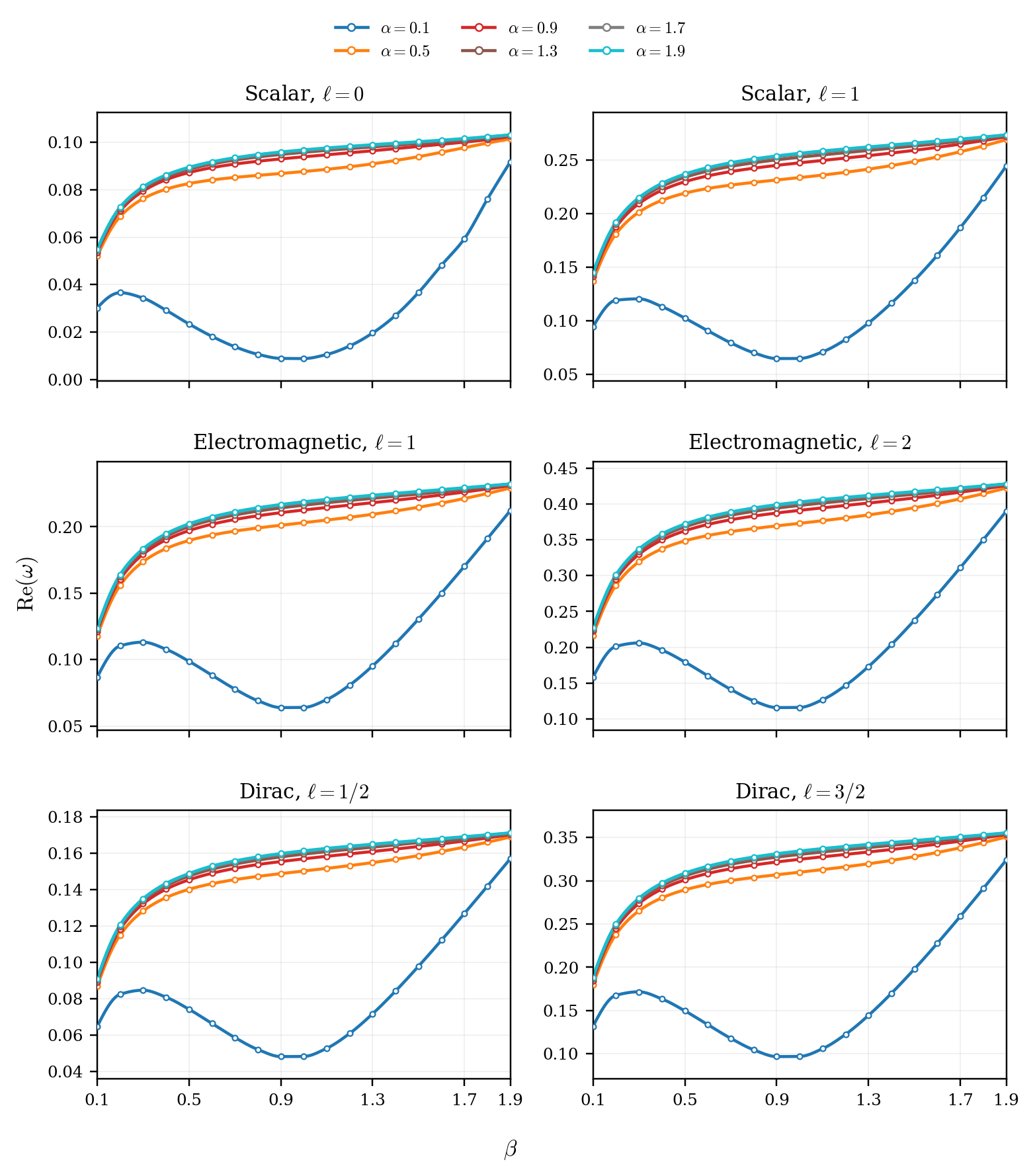}
\caption{Real parts of the fundamental quasinormal frequencies for the dense $(\alpha,\beta)$ scan of six representative sectors at fixed $Q=0$, $M=1$, $\lambda=0.2$, and $c_1=2$. The two scalar panels correspond to $\ell=0$ and $\ell=1$, $\beta$ runs along the horizontal axis, and each curve corresponds to one fixed value $\alpha\in\{0.1,0.5,0.9,1.3,1.7,1.9\}$ using the dense grid $\beta=0.1,0.2,\ldots,1.9$. Markers show the tabulated WKB16 points and the solid lines are interpolations between them.}
\label{fig:alpha_beta_re}
\end{figure*}
\begin{figure*}[t]
\centering
\includegraphics[width=0.98\textwidth]{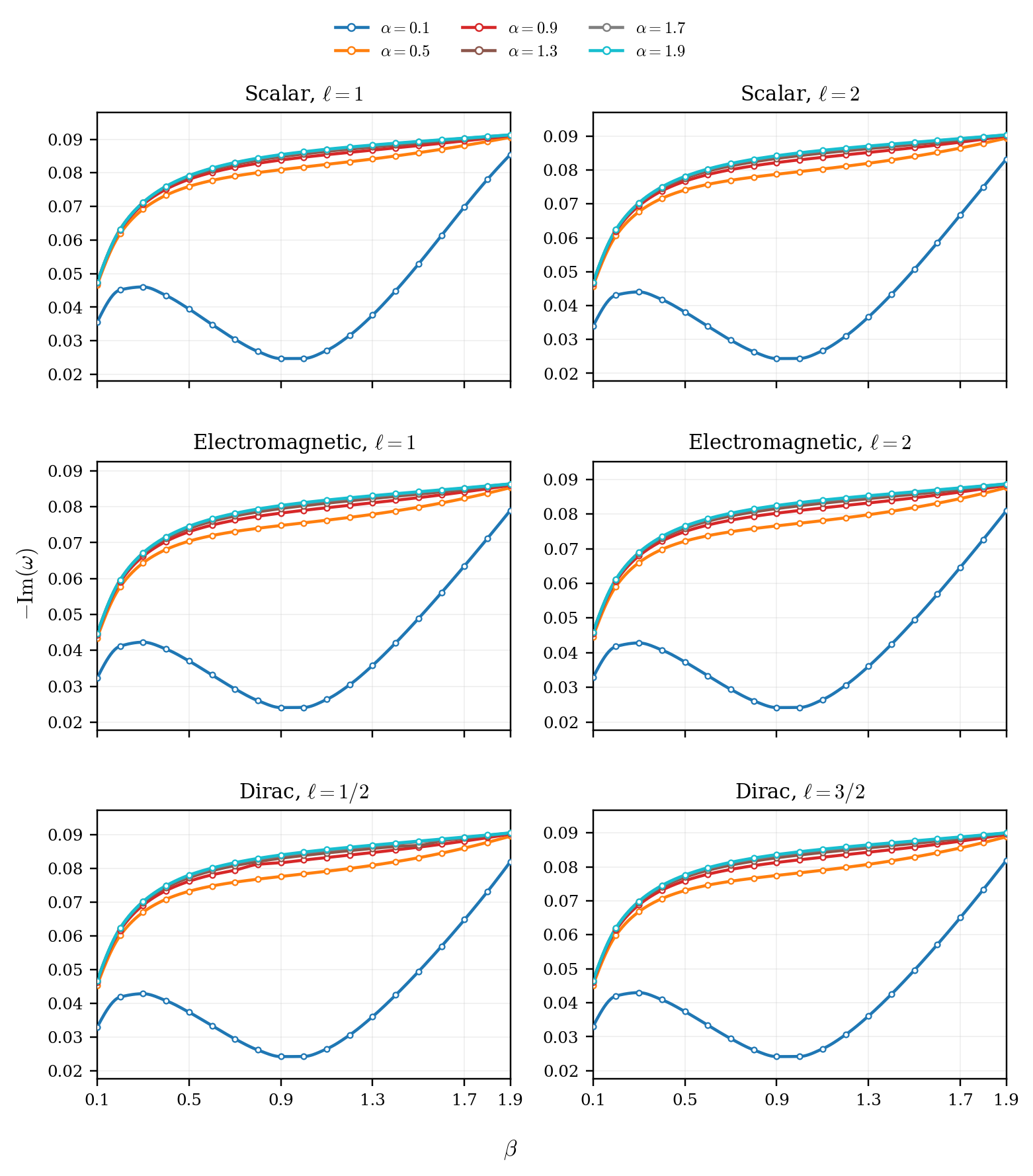}
\caption{Damping rates $-\mathrm{Im}(\omega)$ for the same scan as in Fig.~\ref{fig:alpha_beta_re}, again plotted as functions of $\beta$ for fixed $\alpha$ on the dense grid $\beta=0.1,0.2,\ldots,1.9$. In this figure the two scalar panels correspond to $\ell=1$ and $\ell=2$, while the remaining four panels are unchanged. Markers show the tabulated WKB16 points and the solid lines are interpolations between them.}
\label{fig:alpha_beta_im}
\end{figure*}

\section{Quasinormal Modes}

We will calculate quasinormal modes mainly with the help of the higher order WKB method with Padé approximants. The time-domain integration will be used for an additional check. For the example of the neutral scalar $\ell=1$ mode at $(M,Q,\alpha,\beta,\lambda,c_1)=(1,0,1,1,0.2,2)$, we show that the agreement between the frequency-domain results and the characteristic time-domain integration is excellent; see Fig.~\ref{fig:td-reference-neutral-massless}. 

For clarity, we organize the numerical discussion into three cases. Throughout this section we fix
\begin{equation}
M=1.
\end{equation}
The neutral background is
\begin{equation}
(\alpha,\beta,\lambda,c_1,Q)=(1,1,0.2,2,0),
\end{equation}
for which the two outer horizons are located at
\begin{equation}
(r_h,r_c)\simeq (2.248,17.776).
\end{equation}
All remaining scans vary one parameter family at a time around this point. The numerical accuracy is high: in Table~\ref{tab:qscan_all} the relative difference between the Pad\'e-improved WKB results at orders 16 and 14 never exceeds $0.141\%$, whereas the parameter-induced shifts in the frequencies are at the level of several percent and often tens of percent. Therefore the relative error is evidently tiny and much smaller than the physical effect discussed below.

Figure~\ref{fig:td-reference-neutral-massless} provides the first direct time-domain benchmark for the neutral background $(M,Q,\alpha,\beta,\lambda,c_1)=(1,0,1,1,0.2,2)$ in the scalar $\ell=1$ sector. A Prony fit of the Schwarzschild-like black-hole branch yields $\omega_0=0.249251-0.0849858\,i$ and $\omega_1=0.228648-0.262598\,i$. The corresponding Pad\'e-WKB values are $0.249251-0.084986\,i$ for the fundamental mode (Table~\ref{tab:qscan_all}) and $0.228643-0.262594\,i$ for the first overtone, with the companion WKB14 estimate $0.228643-0.262595\,i$. Thus the fundamental agrees with the WKB result to the displayed digits, while for the first overtone the relative differences are only about $2.2\times 10^{-3}\%$ in the real part and $1.5\times 10^{-3}\%$ in the damping rate. The same profile also resolves the de Sitter branch, for which the extracted frequencies are $\omega_0^{\rm dS}=-0.0525948\,i$ and $\omega_1^{\rm dS}=-0.158721\,i$. For the same asymptotic couplings, the exact empty-de-Sitter formula yields $-iH=-0.052624\,i$ and $-3iH=-0.157871\,i$, so the time-domain damping rates differ by only about $0.055\%$ and $0.54\%$, respectively.

\begin{itemize}
\item \textbf{Case I. Varying the Proca-hair parameter $Q$.} Keeping $(\alpha,\beta,\lambda,c_1)=(1,1,0.2,2)$ fixed, the event horizon persists up to approximately
\begin{equation}
Q_{\max}(M=1)\simeq 1.03.
\end{equation}
We therefore use the discrete set
\begin{equation}
Q=0,\ 0.2,\ 0.4,\ 0.6,\ 0.8,\ 1.0,\ 1.02.
\end{equation}
This scan follows the transition from the ordinary two-horizon sector to the onset of the charged three-horizon regime. Representative roots are
\begin{equation}
\begin{split}
Q=0:&\qquad (r_h,r_c)\simeq (2.248,17.776),\\
Q=0.8:&\qquad (r_h,r_c)\simeq (1.874,17.783),\\
Q=1.0:&\qquad (r_-,r_h,r_c)\simeq (0.758,1.481,17.784),\\
Q=1.02:&\qquad (r_-,r_h,r_c)\simeq (0.941,1.371,17.784).
\end{split}
\end{equation}
The effect on the quasinormal spectrum is clear in all representative sectors collected in Table~\ref{tab:qscan_all}. As $Q$ increases, the damping rates decrease, so the ringing becomes longer lived. Quantitatively, the values of $-\mathrm{Im}(\omega)$ are reduced by roughly $20$--$25\%$ between $Q=0$ and $Q=1.02$. At the same time, the real parts of the $\ell=1$ scalar, electromagnetic, and Dirac modes increase by about $11$--$16\%$. The main exception is the scalar $\ell=0$ mode: its real part first increases up to about $Q\approx0.8$ and then decreases slightly as the three-horizon regime is approached. This behavior is natural because the lowest scalar multipole is directly influenced by the additional curvature term $f'(r)/r$ in Eq.~\eqref{eq:scalar_potential_massless}, whereas the electromagnetic and Dirac sectors are controlled by more uniform barrier structures. The WKB14/WKB16 discrepancies in this table remain tiny throughout and are far below the size of the physical drift with $Q$.

\item \textbf{Case II. Varying $\alpha$ and $\beta$.} To isolate the effect of the asymptotic de Sitter couplings, we keep $Q=0$ and fix $\lambda=0.2$ and $c_1=2$. On this slice, the asymptotically de Sitter black-hole sector exists for $\alpha\gtrsim 0.0922$ and $0<\beta<2$, and in the numerical dataset we sample
\begin{equation}
0.1\leq \alpha\leq 1.9,
\qquad
0.1\leq \beta\leq 1.9,
\label{eq:massless_scan_alpha_beta}
\end{equation}
with $\alpha\in\{0.1,0.5,0.9,1.3,1.7,1.9\}$ and the dense grid $\beta=0.1,0.2,\ldots,1.9$. The geometric deformation is already substantial: varying $\alpha$ from $0.1$ to $1.9$ changes $(r_h,r_c)$ from approximately $(2.887,4.021)$ to $(2.228,25.003)$, whereas varying $\beta$ from $0.1$ to $1.9$ changes $(r_h,r_c)$ from approximately $(3.988,43.504)$ to $(2.133,40.201)$. The figures show that $\beta$ is the parameter that most directly hardens the spectrum. For the practically relevant families with $\alpha\geq 0.5$, increasing $\beta$ increases both ${\rm Re}\,\omega$ and $-\mathrm{Im}(\omega)$ in all six representative sectors displayed in Figs.~\ref{fig:alpha_beta_re} and \ref{fig:alpha_beta_im}. The smallest sampled value $\alpha=0.1$ is exceptional: there the curves develop a shallow minimum near $\beta\sim 0.9$--$1.0$ before turning upward again. By contrast, increasing $\alpha$ at fixed $\beta$ raises the oscillation frequency in all sectors, while its influence on the damping rate is milder and slightly nonmonotonic, most visibly for the scalar $\ell=0$ mode. Thus $\alpha$ mainly rescales the overall frequency level, whereas $\beta$ controls the stronger drift of both the real and imaginary parts.

\item \textbf{Case III. Varying $\lambda$ and $c_1$.} For the remaining couplings we fix $(M,Q,\alpha,\beta)=(1,0,1,1)$ and explore
\begin{equation}
0.05\leq \lambda\leq 0.4,
\qquad
1.5\leq c_1\leq 3.5.
\label{eq:massless_scan_lambda_c1}
\end{equation}
This part of the scan produces the strongest overall softening of the spectrum. Geometrically, $\lambda$ has the largest effect on the size of the static patch, changing $(r_h,r_c)$ from approximately $(2.052,77.954)$ to $(2.719,7.327)$ across the sampled interval, while varying $c_1$ from $1.5$ to $3.5$ changes $(r_h,r_c)$ from approximately $(2.232,25.665)$ to $(2.298,10.728)$. The corresponding spectral effect is systematic in all representative sectors shown in Figs.~\ref{fig:lambda_c1_re} and \ref{fig:lambda_c1_im}: larger $\lambda$ lowers both ${\rm Re}\,\omega$ and $-\mathrm{Im}(\omega)$, and larger $c_1$ amplifies the same tendency. For example, in the scalar $\ell=1$ sector the frequency moves from $0.278020-0.092684 i$ at $(c_1,\lambda)=(1.1,0.1)$ to $0.047048-0.017789 i$ at $(3.5,0.4)$, with analogous behavior in the electromagnetic and Dirac sectors. Hence this scan shows that the combined increase of $\lambda$ and $c_1$ suppresses the oscillatory response most strongly among the parameter families considered here.
\end{itemize}

Taken together, the three cases reveal a clear spin hierarchy. The scalar $\ell=0$ mode is the most model-sensitive sector, because the curvature term in the scalar potential makes it respond more strongly to changes in the background geometry. The electromagnetic sector and the higher-multipole scalar modes provide the cleanest monotonic benchmarks, while the Dirac modes interpolate smoothly between them. In this sense, most of the spin dependence can be traced directly to the structure of the corresponding effective potentials rather than to any ambiguity in the numerical method.

A further issue concerns strong cosmic censorship. For the two-horizon configurations the standard ratio $\beta_{\rm SCC}\equiv\alpha_{\rm gap}/\kappa_-$ is not defined, because there is no inner Cauchy horizon. However, the points $Q=1.0$ and $Q=1.02$ already belong to the charged three-horizon sector, so the SCC diagnostic becomes meaningful there. The displayed data already show that the representative modes become longer lived as this regime is approached, but in the small-black-hole regime the decisive input comes from the exact empty-de Sitter limit. For the same generalized Proca branch, the scalar quasinormal frequencies in the no-black-hole case are known analytically~\cite{Malik:2026dSQNM}:
\begin{equation}
\omega_{n\ell}^{(\pm)}=-iH\left(2n+\ell+\frac{3}{2}\pm\sqrt{\frac{9}{4}-\frac{\mu^2}{H^2}}\right).
\end{equation}
For the massless field they reduce to
\begin{equation}
\omega_{n\ell}^{(1)}=-iH(2n+\ell),
\qquad
\omega_{n\ell}^{(2)}=-iH(2n+\ell+3).
\end{equation}
In particular, the empty-de-Sitter scalar branch contains the static mode $\omega_{00}^{(1)}=0$. Therefore, when a charged black hole is small compared with the de Sitter radius, $r_h/r_c\ll 1$, the corresponding de Sitter-like branch can be arbitrarily weakly damped and naturally controls the spectral gap. For the representative three-horizon point $Q=1.0$ used above one has $(r_-,r_h,r_c)\simeq(0.758,1.481,17.784)$, so $r_h/r_c\simeq 8.33\times 10^{-2}$, and the inner-horizon surface gravity is $\kappa_-\simeq 0.1269$. Hence in the small-black-hole regime one expects $\alpha_{\rm gap}\to0$ for the de Sitter-like branch and therefore $\beta_{\rm SCC}=\alpha_{\rm gap}/\kappa_-\to0$, which automatically satisfies both the smooth-data threshold $1/2$ and the rough-data threshold $1$~\cite{Cardoso:2017soq,Dias:2018etb,KonoplyaZhidenko:2022SCC}. Away from the strictly small-black-hole limit, one must still compute the dominant charged-sector frequencies explicitly, so the present results identify the mechanism responsible for SCC protection in small black holes but do not yet provide a global verdict throughout the full three-horizon domain.

Finally, the quasinormal modes obtained in the present paper can also be used to estimate the grey-body factors through the correspondence developed in Refs.~\cite{Konoplya:2024lir,Konoplya:2024vuj}:
\begin{equation}\label{transmission-eikonal}
\Gamma_{\ell}(\Omega)\equiv |T|^2=
\left(1+e^{2\pi\dfrac{\Omega^2-(\operatorname{Re}\omega_0)^2}{4(\operatorname{Re}\omega_0)(\operatorname{Im}\omega_0)}}\right)^{-1}
+ \mathcal{O}(\ell^{-1}).
\end{equation}
Here $\Gamma_{\ell}(\Omega)$ is the grey-body factor for the partial wave with multipole number $\ell$, $T$ is the transmission amplitude, and $\Omega$ is the real frequency in the scattering problem. The symbols $\operatorname{Re}$ and $\operatorname{Im}$ denote the real and imaginary parts of the quasinormal frequency. The quantity $\omega_0$ is the fundamental mode, while the subleading correction terms contain, in particular, the first overtone $\omega_1$. This correspondence has been tested in a number of works~\cite{Dubinsky:2024vbn,Lutfuoglu:2025ldc,Malik:2025dxn,Bolokhov:2026eqf,Lutfuoglu:2025eik,Malik:2024wvs,Lutfuoglu:2025kqp,Bolokhov:2024otn,Malik:2024cgb,Lutfuoglu:2025mqa,Konoplya:2010vz}. Notice that the correspondence works only with the Schwarzschild branch of modes and cannot be applied to the de Sitter branch \cite{Konoplya:2022gjp}.

\section{Conclusions}

We analyzed massless scalar, electromagnetic, and Dirac perturbations of asymptotically de Sitter black holes in generalized Proca theory on one and the same background geometry. This common setting makes the spin comparison especially transparent: the scalar sector is distinguished by the additional curvature term $f'(r)/r$, the electromagnetic sector is governed by a purely centrifugal barrier, and the Dirac field is described by a pair of isospectral supersymmetric partner potentials. As a result, the scalar $\ell=0$ mode is the most sensitive to changes in the geometry, the electromagnetic sector and higher-multipole scalar modes provide the cleanest monotonic trends, and the Dirac spectrum interpolates smoothly between them.

The parameter scans reveal a consistent spectral pattern. Increasing the Proca-hair parameter $Q$ drives the representative modes toward weaker damping as the geometry approaches the charged three-horizon regime. In the $(\alpha,\beta)$ scan, $\beta$ is the coupling that most directly hardens the spectrum by increasing both ${\rm Re}\,\omega$ and $-\mathrm{Im}(\omega)$ for the physically relevant families, whereas $\alpha$ mainly shifts the overall frequency scale. In the $(c_1,\lambda)$ scan, increasing either coupling softens the ringing, with the combined increase of $\lambda$ and $c_1$ producing the strongest suppression of both oscillation frequency and damping among the parameter families considered here.

The exact empty-de-Sitter scalar spectrum also clarifies the strong-cosmic-censorship interpretation for sufficiently small charged black holes. Because the de Sitter-like branch contains the static mode $\omega_{00}^{(1)}=0$, the corresponding spectral gap can become arbitrarily small compared with the inner-horizon surface gravity, so $\beta_{\rm SCC}$ is expected to tend to zero in the small-black-hole regime. At the same time, the quasinormal frequencies obtained here can be used as input for estimating grey-body factors through the known QNM/grey-body correspondence. For the neutral scalar $\ell=1$ mode at $(M,Q,\alpha,\beta,\lambda,c_1)=(1,0,1,1,0.2,2)$, the agreement with the time-domain integration is excellent; see Fig.~\ref{fig:td-reference-neutral-massless}. The remaining next steps are therefore to determine the dominant de Sitter-like modes explicitly throughout the full charged three-horizon sector and to compute the associated grey-body factors directly for the scalar, electromagnetic, and Dirac fields.


\clearpage
\bibliographystyle{apsrev4-1}
\bibliography{ProcaMassiveBH,referencesProca}

\end{document}